\documentclass[12pt]{article}
\usepackage{amsmath}
\usepackage{amssymb}
\usepackage{amsfonts}
\newcommand{\ba}{\begin{array}{l}}
\newcommand{\ea}{\end{array}}
\newcommand{\beq}{\begin{equation}}
\newcommand{\eeq}{\end{equation}}
\newcommand{\bea}{\begin{eqnarray}}
\newcommand{\eea}{\end{eqnarray}}
\usepackage{epsfig}
\usepackage{graphicx}
\usepackage{color}

\definecolor{dyellow}{rgb}{1.,0.8,.0}
\definecolor{myblue}{rgb}{.1,.1,.7}
\definecolor{dcyan}{rgb}{.0,.6,.6}
\definecolor{dmagenta}{rgb}{0.6,0.0,0.6}
\definecolor{brown}{rgb}{0.6,0.2,0.}
\definecolor{darkblue}{rgb}{.0,.0,0.5}
\definecolor{darkred}{rgb}{0.75,0.0,0.0}
\definecolor{orange}{rgb}{1.,.6,.0}
\definecolor{dorange}{rgb}{0.8,.4,.0}
\definecolor{darkgreen}{rgb}{0.0,0.6,0.0}
\definecolor{purple}{rgb}{.4,.0,.4}

\def\bc{\begin{center}}
\def\ec{\end{center}}
\def\be{\begin{eqnarray}}
\def\ee{\end{eqnarray}}

\newcommand{\omits}[1]{}

\begin{document}
\begin{center}
{\Large \bf {  Modified Newton's gravity in Finsler Space as a possible alternative to dark matter hypothesis}}\\
  \vspace*{1cm}
Zhe Chang \footnote{changz@mail.ihep.ac.cn} and  Xin
Li \footnote{lixin@mail.ihep.ac.cn}\\
\vspace*{0.2cm} {\sl Institute of High Energy Physics,
Chinese Academy of Sciences\\
P. O. Box 918(4), 100049 Beijing, China}\\

\bigskip

\end{center}
\vspace*{2.5cm}

%
\begin{abstract}
A modified Newton's gravity is obtained as the weak field
approximation of the Einstein's equation in Finsler space. It is
found that a specified Finsler structure makes the modified Newton's
gravity equivalent to the modified Newtonian dynamics (MOND). In the
framework of Finsler geometry, the flat rotation curves of spiral
galaxies can be deduced naturally without invoking dark matter.
 \vspace{1cm}
\begin{flushleft}
\end{flushleft}
\end{abstract}

\newpage

There are a great variety of observations which show that  the
rotational velocity curves of all spiral galaxies tend to some
constant values\cite{Trimble}. These include the Oort discrepancy in
the disk of the Milky Way\cite{Bahcall}, the velocity dispersions of
dwarf Spheroidal galaxies\cite{Vogt}, and the flat rotation curves
of spiral galaxies\cite{Rubin}. These facts
 violate sharply the prediction of Newtonian dynamics or Newton's gravity.

The most widely adopted way to resolve these difficulties is the
dark matter hypothesis. It is assumed that all visible stars are
surrounded by massive nonluminous matters. Though it explains the
flat rotation curves of spiral galaxies, the hypothesis has its own
weakness. No theory predicts these matters, and they behave in such
ad hoc way. There are a lot of possible candidates of dark matter
(such as axion, neutrino {\it et al}), but none of them
satisfactory. Up to now, all of them either undetected or excluded
by observation.

Because of these troubles induced by dark matter, some models have
been built for alternative of the dark matter hypothesis. Their main
ideas are to assume that the Newtonian gravity or Newton's dynamics
is invalid in galactic scale. In particular,  two models explain
well the flat rotational curves successfully without invoking dark
matter. One is higher-order gravitational theory\cite{Xu}. The
gravitational potential was supposed\cite{Sanders}
 of the Yukawa form, \be\label{Yukawa}
\varphi(r)=-\frac{GM}{r(1+\alpha)}\left(1+\alpha e^{r/r_0}\right).
\ee Another is the famous MOND\cite{Milgrom}. It assumed that the
Newtonian dynamics does not hold in galactic scale. The particular
form of MOND is given as
 \begin{equation}
 \begin{array}{l}
  m\mu\left(\displaystyle\frac{a}{a_0}\right)\mathbf{a}=\mathbf{F},\\[0.4cm]
 \displaystyle\lim_{x\gg1}\mu(x)=1,~~~\lim_{x\ll1}\mu(x)= x,
 \end{array}
 \end{equation}
where $a_0$ is at the order of $10^{-8}$ cm/s$^2$. At beginning, as
a phenomenological model, MOND explains well the flat rotation
curves with a simple formula and a new parameter. In particular, it
deduce naturally a well-known global scaling relation for spiral
galaxies, the Tully-Fisher relation\cite{TF}. By introducing several
scalar, vector and tensor fields, Bekenstein\cite{Bekenstein}
rewrote the MOND into a covariant formulation. He showed that the
MOND satisfies all four classical tests on Einstein's general
relativity in Solar system.

These models seem appealing in theoretical interest and fit well the
empirical data of the flat rotation curves. However, it has been
pointed out that solving dark matter problem by means of metric
theories is too difficult. The Yukawa term(\ref{Yukawa}) is mediated
by a spin 1 vector particle\cite{Sanders}. Such a
contribution\cite{Zhytnikov} may be proportional to baryonic charge.
The baryonic charge varies from one body to another, this violates
the weak equivalence principle. MOND, as a very successful
phenomenological model, still face problems yet. Reproducing the
MOND force law\cite{Soussa} requires any completely stable,
metric-based theory of gravity to become conformally invariant in
the weak field limit, and the prospects for a formulation with a
very weak instability.

In this Letter, we present another possible alternative to the dark
matter hypothesis. The major property of the flat rotational curves
is the velocity of a particle in a circular orbit around a finite
spiral galaxy becomes independent of the radius of the orbit at
large radii. This empirical fact makes us consider the Finsler
geometry. In Finsler geometry, the intrinsic curve is not only the
function of position but also the function of velocity. And in the
framework of Finsler geometry, the four-velocity vector is treated
as independent variable\cite{Chang}.  The explicit DISIM$_b(2)$
invariant Finslerian line element and the respective Lie algebra was
first proposed thirty years
ago\cite{Bogoslovsky2,Bogoslovsky3,Gibbons}.  Finsler geometry is a
natural and fundamental generalization of Riemann geometry.

The gravity in Finsler space has been investigated for a long
time\cite{Takano,Ikeda,Tavakol, Bogoslovsky1}. In this Letter, we
begin with the field equation in Berwald space-a special space in
Finsler space. The gravitational field equation in Berwald-Finsler
space has been written down explicitly\cite{Lixin}, \be\label{field
equation of Berwald}
\left[Ric_{\mu\nu}-\frac{1}{2}g_{\mu\nu}S\right]+\left\{\frac{1}{2}
B^{~\alpha}_{\alpha~\mu\nu}+B^{~\alpha}_{\mu~\nu\alpha}\right\}=8\pi
G T_{\mu\nu}. \ee We will study the weak field approximation of the
field equation (\ref{field equation of Berwald}). Results obtained
should be compared with Newton's gravity. We wish the modified
Newton's gravity got as the weak field approximation of the field
equation (\ref{field equation of Berwald}) in Finsler space
describes the odd behavior of the rotation curves of spiral
galaxies.

Before dealing with the field equation, we introduce some basic
notation of Finsler geometry\cite{Book by Bao}. Denote by $T_xM$ the
tangent space at $x\in M$, and by $TM$ the tangent bundle of $M$.
Each element of $TM$ has the form $(x, y)$, where $x\in M$ and $y\in
T_xM$. The natural projection $\pi : TM\rightarrow M$ is given by
$\pi(x, y)\equiv x$. A Finsler structure of $M$ is a function\be F :
TM\rightarrow[0,\infty)\nonumber \ee with the following
properties:\\
(i) Regularity: F is $C^\infty$ on the entire slit tangent bundle
$TM\backslash0$.\\
(ii) Positive homogeneity : $F(x, \lambda y)=\lambda F(x,
y)$ for all $\lambda>0$.\\
(iii) Strong convexity: The $n\times n$ Hessian matrix\be
g_{\mu\nu}\equiv\frac{\partial}{\partial
y^\mu}\frac{\partial}{\partial
y^\nu}\left(\frac{1}{2}F^2\right)\nonumber \ee is positive-definite
at every point of $TM\backslash0$.

Finsler geometry has its genesis in integrals of the form
\be\label{integral length} \int^r_s
F\left(x^1,\cdots,x^n;\frac{dx^1}{dt},\cdots,\frac{dx^n}{dt}\right)dt.\ee

Throughout the Letter, the lowering and raising of indices are
carried out by the fundamental tensor $g_{\mu\nu}$ defined above,
and its inverse $g^{\mu\nu}$.

In Finsler manifold, there exist a unique linear connection~-~the
Chern connection\cite{Chern}. It is torsion freeness and
metric-compatibility,
 \be
 \Gamma^{\alpha}_{\mu\nu}=\gamma^{\alpha}_{\mu\nu}-g^{\alpha\lambda}\left(A_{\lambda\mu\beta}\frac{N^\beta_\nu}{F}-A_{\mu\nu\beta}\frac{N^\beta_\lambda}{F}+A_{\nu\lambda\beta}\frac{N^\beta_\mu}{F}\right),
 \ee
 where $\gamma^{\alpha}_{\mu\nu}$ is the formal Christoffel symbols of the
second kind with the same form of Riemannian connection, $N^\mu_\nu$
is defined as
$N^\mu_\nu\equiv\gamma^\mu_{\nu\alpha}y^\alpha-A^\mu_{\nu\lambda}\gamma^\lambda_{\alpha\beta}y^\alpha
y^\beta$
 and $A_{\lambda\mu\nu}\equiv\frac{\partial}{\partial y^\lambda}\frac{\partial}{\partial y^\mu}\frac{\partial}{\partial y^\nu}\frac{F}{4}(F^2)$ is the
 Cartan tensor (regarded as a measurement of deviation from Riemannian Manifold).
The curvature of Berwald~-~Finsler space is given as \be
R^{~\lambda}_{\kappa~\mu\nu}&=&\frac{\partial
\Gamma^\lambda_{\kappa\nu}}{\partial x^\mu}-\frac{\partial
\Gamma^\lambda_{\kappa\mu}}{\partial
x^\nu}+\Gamma^\lambda_{\alpha\mu}\Gamma^\alpha_{\kappa\nu}-\Gamma^\lambda_{\alpha\nu}\Gamma^\alpha_{\kappa\mu}.\ee
The Ricci tensor on Finsler manifold was first introduced by
Akbar-Zadeh\cite{Akbar}. In Berwald-Finsler space, it reduces to \be
Ric_{\mu\nu}=\frac{1}{2}(R^{~\alpha}_{\mu~\alpha\nu}+R^{~\alpha}_{\nu~\alpha\mu}).\ee
It is manifestly symmetric and covariant. Apparently the Ricci
tensor will reduce to the Riemann-Ricci tensor if the Cartan tensor
vanish identically. The trace of the Ricci tensor gives the scalar
curvature $S\equiv g^{\mu\nu}Ric_{\mu\nu}$. We starts from the
second Bianchi identities on Berwald-Finsler space \be\label{Bianchi
on Riemann} R^{~\alpha}_{\mu~\lambda\nu|\beta}+R^{~\alpha}_{\mu~\nu
\beta|\lambda}+R^{~\alpha}_{\mu~\beta\lambda|\nu}=0,\ee where the
$|$ means the covariant derivative. The metric-compatibility \be
g_{\mu\nu|\alpha}=0~~~~\mathrm{and}~~~~g^{\mu\nu}_{~~|\alpha}=0,\ee
and contraction of (\ref{Bianchi on Riemann}) with $g^{\mu\beta}$
gives that \be R^{\mu
\alpha}_{~~\lambda\nu|\mu}+R^{\mu\alpha}_{~\nu\mu|\lambda}+R^{\mu\alpha}_{~\mu\lambda|\nu}=0.
\ee Lowering the index $\alpha$ and contracting with
$g^{\alpha\lambda}$,  we obtain \be
\left[Ric_{\mu\nu}-\frac{1}{2}g_{\mu\nu}S\right]_{|\mu}+\left\{\frac{1}{2}B^{~\alpha}_{\alpha~\mu\nu}+B^{~\alpha}_{\mu~\nu\alpha}\right\}_{|\mu}=0
,\ee where \be
B_{\mu\nu\alpha\beta}=-A_{\mu\nu\lambda}R^{~\lambda}_{\theta~\alpha\beta}y^\theta/F.\ee
 Thus, the counterpart of the Einstein's field equation on
 Berwald~-~Finsler space  takes the form
 \be
\left[Ric_{\mu\nu}-\frac{1}{2}g_{\mu\nu}S\right]+\left\{\frac{1}{2}B^{~\alpha}_{\alpha~\mu\nu}+B^{~\alpha}_{\mu~\nu\alpha}\right\}=8\pi
G T_{\mu\nu}. \ee

To get a modified Newton's gravity, we consider a particle moving
slowly in a weak stationary gravitational field\cite{Weinberg}. We
suppose that the metric is close to the locally Minkowskian metric,

 \begin{equation}\label{metric}
 \begin{array}{l}
  g_{\mu\nu}(x,y)=\eta_{\mu\nu}(y)+h_{\mu\nu}(x,y), \\[0.2cm]
\eta_{\mu\nu}(y)=f(y)*{\rm diag}\{1,-1,-1,-1\}~,
\end{array}\end{equation}
where $|h_{\mu\nu}|\ll1$.

The field is stationary and all time derivatives of $g_{\mu\nu}$
vanish. The Cartan tensor is symmetric in its three indices. To
first order in $h_{\mu\nu}$, we have \be
\Gamma^i_{00}=\frac{1}{2f}\frac{\partial h_{00}}{\partial
x^i}-\frac{1}{4f^2}\frac{\partial f}{\partial y^0}y^0\frac{\partial
h_{00}}{\partial x^i}, \ee and \be Ric_{00}=\frac{\partial
\Gamma^i_{00}}{\partial x^i}. \ee All particles here are moving
slowly. The velocity $y^0$ equal to the speed of light approximately
and can be regarded as constant. Then, the connection
$\Gamma^i_{00}$ reduces as \be
\Gamma^i_{00}=\frac{1}{2f}\frac{\partial h_{00}}{\partial x^i}. \ee
From the definition of $B_{\mu\nu\alpha\beta}$, to first order in
$h_{\mu\nu}$, we know that all components of $B$ vanish. By making
use of the non-relativistic perfect fluid approximation, we obtain
the non vanished component of energy-momentum tensor $T_{00}=\rho
g_{00}$. Here $\rho$ is the proper energy density. After the above
manipulations, the velocity four-vector reduces to the normal
velocity $v$, which is defined in the manifold $M$ and function of
distance. In the above approximations, the field equation(\ref{field
equation of Berwald}) reduces to
 \be\label{field equation1} \frac{\partial}{\partial x_i}\left(\frac{1}{2f(v)}\frac{\partial h_{00}}{\partial x^i}\right)=4\pi G\rho
f(v). \ee
 The geodesic equation in Finsler space is given as\cite{Book by
 Bao}
 \be\label{geodesic} \frac{d^2 x^\lambda}{d
 \tau^2}+\gamma^\lambda_{\mu\nu}\frac{dx^\mu}{d\tau}\frac{dx^\nu}{d\tau}=0.
 \ee
 Though the form of the geodesic equation is the same with the
  one in Riemann geometry, but one should notice that
 the connection $\gamma^\lambda_{\mu\nu}$ here depends both on coordinates and velocities. We
 will show the physical meaning of the velocity dependence in an
 indirect way by solving the geodesic equation in weak field
 approximation.
In the approximation of moving slowly and weak field,
 the geodesic equation(\ref{geodesic}) reduces to
 \be\label{geodesic equation} \frac{d^2t}{d\tau^2}&=&0,\nonumber\\
 \frac{d^2x^i}{d\tau^2}&=&-\frac{1}{2f}\frac{\partial h_{00}}{\partial
 x^i}\left(\frac{dt}{d\tau}\right)^2. \ee
The solution of first equation in (\ref{geodesic equation}) is
$dt/d\tau=const.$. Dividing the second equation in (\ref{geodesic
equation}) by $(dt/d\tau)^2$, we obtain
 \be\label{potential} \frac{d^2x^i}{dt^2}=-\frac{1}{2f}\frac{\partial h_{00}}{\partial
 x^i}\equiv-\frac{\partial \varphi}{\partial x^i},\ee
 where $\varphi$ is the gravitational potential. Substituting the
 relation (\ref{potential}) into equation (\ref{field
 equation1}), we get
 \be\label{poisson equation} \nabla^2\varphi=4\pi G\rho f(v). \ee
The velocity $v$ is a function of coordinate. In spherical
coordinate, we can rewrite $f(v)$ into $\rho_{geo}(r)$. The
integration of (\ref{poisson equation}) gives \be\label{modified
gravity} \nabla\varphi(r)=\frac{G}{r^2}\int \rho(r)\rho_{geo}(r)
dV=\frac{G\mathcal{M}}{r^2} .\ee One can see clearly that the
effective mass $\mathcal{M}$ is different from the baryonic mass. It
is enlarged by the factor $\rho_{geo}(r)$ due to Finsler geometrical
effect.

We suppose the energy density of baryons is constant for
convenience. Denotes $M$ the mass of visible matter.

We would like limit the metric (\ref{metric}) to be the form

\begin{equation}
\begin{array}{l}
\displaystyle f(\xi)=\frac{\frac{5}{6}\frac{3a_0}{4\pi
G\rho\xi}+1}{\sqrt{\frac{3a_0}{4\pi
G\rho\xi}+1}}~,\\[1cm]
v=\left(\frac{4}{3}\pi Ga_0\rho\xi^3+\left(\frac{4}{3}\pi
G\rho\right)^2\xi^4\right)^{1/4}~.
\end{array}\end{equation}
Here $a_0$ is the deformation parameter of Finsler geometry. It is
the measurement of deviation from Riemann geometry. The deformation
of Finsler space may have cosmological significance. One wishes
naturally the deformation parameter relates with the cosmological
constant $\Lambda$. In fact, Milgrom observed\cite{Milgrom} that
$2\pi a_0\approx c(\sqrt{\Lambda/3})$. Eqs (\ref{poisson equation})
and (\ref{modified gravity}) give the geometrical factor of the
density of baryons,
 \be
\rho_{geo}(r)=\frac{\frac{5}{6}+\frac{GM}{r^2a_0}}{\sqrt{(\frac{GM}{r^2a_0})^2+\frac{GM}{r^2a_0}}}.
\ee
 In the zero limit of the deformation parameter, familiar results on
 Riemann geometry are recovered.
Thus, the acceleration $a$ is expressed as \be
a=\nabla\varphi(r)=\frac{GM}{r^2}\sqrt{\frac{r^2+GM/a_0}{GM/a_0}}.\ee
We can rewrite it in a compact form,  \be
a=\frac{GM}{r^2}\nu\left(\frac{GM}{r^2a_0}\right),\ee where \be
\nu(x)=\sqrt{\frac{1+x}{x}}\approx\left\{ 1 ~~~~~~~{\rm for}~~ x\gg
1\atop x^{-1/2} ~~{\rm for}~~ x\ll 1 \right. .\ee It is obvious that
the above formula is just the MOND \cite{MOND}. The MOND reduces to
Newton's gravity when $x\gg1$. Meanwhile, in our model the metric
$\eta_{\mu\nu}$ returns to Minkowskian metric.

Persic, Salucci and Stel analyzed about 1100 rotational curves of
spiral galaxies\cite{Persic}. They gave a universal formula for the
rotation curves, called {\it Universal Rotation Curves} (URC),
 \be
 V_{URC}(r)=V(r_{opt})\bigg[\left(0.72+0.44\log\frac{L}{L_{\ast}}\right)\frac{1.97x^{1.22}}{(x^2+0.78^2)^{1.43}}\nonumber\\
 +1.6e^{-0.4(L/L_{\ast})}\frac{x^2}{x^2+1.5^2(L/L_{\ast})^{0.4}}\bigg]^{1/2}
 {\rm km/s}, \ee
 where $r_{opt}$ is the radius encircling 83\% of the integrated
light and $x=r/r_{opt}$. The luminous matter in spiral galaxies is
composed by spheroidal bugle and an extended thin exponential disk.
The spheroidal bugle has density distribution of the form
$I(r)=I_0e^{-7.67(r/r_e)^{1/4}}$(where $r_e$ is the half-light
radius)\cite{Vaucouleurs}. The thin exponential disk has density
distribution of the form $I(r)=I_0e^{-r\alpha}$($\alpha^{-1}$ is the
disk scale-length)\cite{Freeman}. Phenomenological similarities
between the URC and MOND have been investigated in detail by looking
for properties predicted in one framework that are also reproducible
in the other one\cite{Gentile}. We wish a general Finslerian metric
be found to reproduced the URC naturally. In fact, many works about
dark matter have been done, and the density profile of dark matter
halos surrounding galaxies have been given. The Finsler geometrical
factor is determined by \be \rho_{geo}=1+\rho_{DM}/\rho .\ee

\bigskip

\centerline{\large\bf Acknowledgements} \vspace{0.5cm}
 We would like to thank T. Chen,
P. Wang, N. Wu and Y. Yu for useful discussion. The work was
supported by the NSF of China under Grant NO. 10575106.


\end{document}